\documentclass{lmcs} 
\pdfoutput=1

\usepackage{lastpage}
\lmcsdoi{15}{3}{22}
\lmcsheading{}{\pageref{LastPage}}{}{}%
{Jul.~04,~2018}{Aug.~28,~2019}{}


\usepackage{hyperref}
\usepackage{amssymb}
\usepackage{amsmath}
\usepackage{textcomp}

\usepackage{color}
\usepackage{comment}

\usepackage{graphicx}
\usepackage{ifpdf}

\newcommand{\bra}[1]{\langle #1|}
\newcommand{\ket}[1]{|#1\rangle}

\newcommand{\cent}[0]{\mbox{\textcent}}
\newcommand{\dollar}[0]{\$}

\newcommand{\twin}{\mathtt{TWIN}}
\newcommand{\usqr}{\mathtt{USQUARE}}

\newcommand{\mypar}[1]{ \left( #1 \right) }
\newcommand{\mymatrix}[2]{ \mypar{ \begin{array}{#1} #2 \end{array} } }
\newcommand{\myvec}[1]{ \mymatrix{c}{#1} }

\begin{document}

\title{Alternating, private alternating, and quantum alternating realtime automata}\titlecomment{A preliminary version appeared as ``H.~G{\"{o}}kalp Demirci, Mika Hirvensalo, Klaus Reinhardt, A.~C.~Cem Say, and
  Abuzer Yakary{\i}lmaz.
\newblock Classical and quantum realtime alternating automata.
\newblock In {\em NCMA}, volume 304, pages 101--114. {\"{O}}sterreichische
  Computer Gesellschaft, 2014.'' \cite{DHRSY14}. The arXiv number is 1407.0334.}

\author[G. Dem\.{i}rc\.{i}]{G\"okalp Dem\.{i}rc\.{i}}
\address{University of Chicago, Department of Computer Science, Chicago, IL 60637 USA}
\email{demirci@cs.uchicago.edu}

\author[M. Hirvensalo]{Mika Hirvensalo}
\address{Department of Mathematics and Statistics, University of Turku, FI-20014 Turku Finland}
\email{mikhirve@utu.fi}

\author[K. Reinhardt]{Klaus Reinhardt}
\address{Institut f\"{u}r Informatik, University of Halle-Wittenberg, D-06120 Halle Germany}
\email{klaus.reinhardt@informatik.uni-halle.de}

\author[A. C. C. Say]{A. C. Cem Say}
\address{Bo\u{g}azi\c{c}i University, Department of Computer Engineering, Bebek 34342 \.{I}stanbul, Turkey}
\email{say@boun.edu.tr}

\author[A. Yakary{\i}lmaz]{Abuzer Yakary{\i}lmaz}
\address{University of Latvia, Center for Quantum Computer Science, Raina bulv. 19, R\={\i}ga, LV-1586, Latvia}
\email{abuzer@lu.lv}

\begin{abstract}
We present new results on realtime alternating, private alternating, and quantum alternating automaton models. Firstly, we show that the emptiness problem for alternating one-counter automata on unary alphabets is undecidable. Then, we present two equivalent definitions of realtime private alternating finite automata (PAFAs). We show that the emptiness problem is undecidable for PAFAs. Furthermore, PAFAs can recognize some nonregular unary languages, including the unary squares language, which seems to be difficult even for some classical counter automata with two-way input. Regarding quantum finite automata (QFAs), we show that the emptiness problem is undecidable both for universal QFAs on general alphabets, and for alternating QFAs with two alternations on unary alphabets. On the other hand, the same problem is decidable for nondeterministic QFAs on general alphabets. We also show that the unary squares language is recognized by alternating QFAs with two alternations.
\end{abstract}

\maketitle

\section{Introduction}

Alternation is a generalization of nondeterminism \cite{CKS81}. Although alternating finite automata (AFAs) can   only recognize regular languages, even when they can use a two-way input head,  they can be more powerful when given other resources. For example, AFAs augmented with a counter (A1CAs) can recognize some unary nonregular languages \cite{RY14A,BGRY16}, whereas their nondeterministic counterparts cannot recognize any unary nonregular language, even when allowed to pause the input head indefinitely on a symbol, and given a stack instead of a counter, which upgrades them to one-way nondeterministic pushdown automata (PDAs) \cite{GR62}. 

It is a well-known fact that the emptiness problem (which asks whether there exists an input that is accepted by a given machine)  is decidable,  but the universality problem (which asks whether all possible inputs are accepted) is undecidable for one-way PDAs. Therefore, the emptiness problem is also undecidable for \textit{universal} pushdown automata, i.e. those with universal branchings, which accept an input only if all computational paths end in accept states. This  result can also be obtained for the realtime version of this model by replacing the stack with a counter, obtaining a A1CA with a single alternation having only universal states. But obtaining  the same result is not trivial for A1CAs with unary input alphabets. In this paper, we prove that the emptiness problem for A1CAs on unary alphabets is undecidable.

Private alternation is a generalization of alternation \cite{Rei79,PR79,Rei84}, which is usually modelled as a game between the existential and universal players (see also \cite{Co89}), where some computational resources (tape heads, working memories, etc) are private to the universal player. Such privacy can increase the computational power of models. For example, if the input head is private, then one-way private AFAs can recognize any alternating linear-space language ($ \mathsf{ASPACE(n)} = \mathsf{DTIME(2^n)} $) by simulating linear-space alternating Turing machines (TMs) on the given inputs. Note that the automaton runs forever in some useless computational paths during this simulation. 

In this paper, we present two  equivalent definitions of realtime versions of private AFAs (PAFAs). The first definition is given as a generalization of AFAs, and the second definition is given based on interactive proof systems. We then show that realtime PAFAs can recognize some nonregular unary languages including the unary squares language, which seems to be difficult even for some classical counter automata with two-way input \cite{IJTW93,Pet94,DBY14A}. We also show that the emptiness problem for PAFAs is undecidable. 

The concept of quantum alternation was introduced   and shown to be very powerful recently \cite{Yak13A,Yak16A}: One-way alternating quantum finite automata (QFAs) can simulate the computation of any Turing machine  on a given input, and so they can recognize any recursively enumerable language. Similar to private AFAs, the automaton runs forever in some useless computational paths during this simulation. In this paper, we focus on their realtime counterparts, and show that  the emptiness problem is undecidable for both universal QFAs on general alphabets and alternating QFAs with two alternations on unary alphabets, but decidable for nondeterministic QFAs. Moreover, we show that the unary squares language is recognized by alternating QFAs with two alternations.

Throughout the paper, $\Sigma$ denotes the input alphabet not containing the end-markers ($\cent$ and $\dollar$) and $\tilde{\Sigma} = \Sigma \cup \{\cent,\dollar\}$. Any given input $w \in \Sigma$ is read as $ \cent  w \dollar$. All alternating realtime models can spend at most two steps on each input symbol, so that they can make both existential and universal choices on the same symbol. For any given set $ S $, $ \mathcal{P}(S) $ is the power set of $ S $.

We classify the results with respect to models; alternation, private alternation, and quantum alternation. We devote a section to each of them, and provide the required background within the sections.

A preliminary version of this paper was presented in NCMA 2014 \cite{DBY14A}. We rewrite the section on private alternation (Section \ref{sec:private-alternation}), extend the proof of Theorem \ref{thm:AQFA-usquare}, and also revise the text.

\section{Alternation}

The alternating Turing machine and its certain variants were introduced in \cite{Cha81}. Here we present our definition of an alternating finite automaton operating in realtime mode. 
A realtime alternating finite automaton is (AFA) a 5-tuple 
\[
	A = (S,\Sigma,\delta,s_1,s_a),
\]
where
\begin{itemize}
	\item $S$ is the finite set of states, composed of existential and universal states,
	\item $s_1$ is the initial state and $s_a$ is the accepting state, and,
	\item $\delta : S \times \tilde{\Sigma} \rightarrow \mathcal{P}(S) $ is the transition function.
\end{itemize}
The automaton spends two steps on each symbol.  When $A$ is in state $s \in S$ and reads  $\sigma \in \tilde{\Sigma}$, it switches to a set of states $S' \subseteq S $, where $\delta(s,\sigma) = S' $. The automaton gives the decision of acceptance if it ends the computation in state $s_a$ (after reading $\dollar$). The computation of an AFA can be shown as a tree such that each node represents a state, and the edges represent the transitions. We associate each inner node with either a ``$\vee$'' or a ``$\wedge$'', depending on whether its corresponding state is existential or universal, respectively. When evaluating the tree, each leaf in which the input is accepted (resp., rejected) takes the value of ``true'' (resp., ``false''). The value of the root can be evaluated from bottom to top: Any existential inner node (assigned a ``$\vee$'') takes the value of ``true'' if at least one of its children has taken the value of ``true'' and any universal inner node (assigned a ``$\wedge$'') takes the value of ``true'' if all of its children have taken the value of ``true''. The input is accepted if and only if the root takes the value of ``true''.

A realtime alternating one-counter automaton (A1CA) is a (realtime) AFA augmented with an integer counter. In each step, the automaton additionally can test  whether the counter value is zero or not, and then it adds a value from  $ \{ -1,0,1 \} $ to the counter in each branch. The transition function is formally defined as  $ \delta: S \times \tilde{\Sigma} \times \{ 0,\pm \} \rightarrow \mathcal{P} ( S \times \{-1,0,1\} )  $, where $ 0 $ and $ \pm $ represent whether the value of the counter is zero or nonzero, respectively.

\begin{thm}
	The emptiness problem for unary A1CAs is undecidable.
\end{thm}
\begin{proof}
	For a deterministic Turing machine $M$, we can construct a realtime alternating counter automaton $A$ which accepts the unary input $u^{2n}$ if and only if $M$ starting on an empty tape halts in exactly $n$ steps.
To solve the emptiness problem for $A$ is then to solve whether $M$ halts on the empty string, a problem whose  undecidability is well known.
The proof idea is similar to Theorem 3.4 in \cite{CKS81}, showing how an alternating Turing machine can mimic deterministic $T(n)$-time computation in $O(\log T(n))$ space by utilizing backwards simulation.

Without loss of generality, we can assume that the tape of $M$ is semi-infinite, cells numbered with nonnegative integers, and that there is always a special starting symbol $\triangleright$ in cell number $0$. The machine never attempts to overwrite $\triangleright$. Moreover, by introducing another left endmarker, we may assume that the final step (if $M$ halts) is the first time $M$ returns to head position $C=0$, and halts in a unique halting state $q_f$.

The simulation uses the input word $u^{2n}$ as a clock measuring the steps of computation, and the counter to store a pointer to a position in a configuration of the  simulated machine. By the aforementioned assumptions, the counter value will start at $C=0$, remain nonnegative, and return to value $C=0$ only if the computation halts. The starting configuration of $M$ is numbered  $0$, the final one is numbered $n$, and the backwards simulation means that the configuration numbered $n-i$ is simulated when reading symbols $2i$ and $2i+1$ of the input word $u^{2n}$. 

We will outline an alternating algorithm for such a backwards simulation of a given TM $M$, and leave the actual formalization of a realtime A1CA embodying this algorithm to the reader. Since $M$ is deterministic, the content of position $C$ of its $t$-th configuration depends only on the contents of positions $C-1$, $C$, and $C+1$ of the $(t-1)$-st configuration. 
These contents may be just a symbol of the alphabet of $M$, or a state-symbol pair from $M$ for the case that $M$ is in the indicated state and the input head is at this position
(this can only be the case for one of the three positions).
Hence there is a finite partial function $s^{(t+1)}_{C}=\rm{Next}_M(s^{(t)}_{C-1},s^{(t)}_{C},s^{(t)}_{C+1})$, where the superscript refers to configuration number, and the subscript to the position of the content.

The automaton $A$ begins in a state representing the final Turing machine configuration: State $q_f$ paired with the scanned symbol $\triangleright$ in memory, and counter $0$ representing this final head position.
The task of automaton $A$ is to check whether such a configuration can be obtained by a computation of Turing machine $M$ in $n$ steps.

In each stage of the checking, automaton $A$ keeps the configuration contents $s$ in its memory, and its target is to check whether $s=s^{(t)}_{C}$. It reads one input symbol and (existentially) guesses a combination $(s_{-1},s_0,s_1)$ with $s=\rm{Next}_M(s_{-1},s_0,s_1)$.
Then $A$ reads another input symbol and universally branches into configurations where $C$ is replaced with $C-1$, $C$, and $C+1$, respectively, and enters  the checking stage again to verify whether $s^{(t-1)}_{C+d}=s_d$ for $d\in\{-1,0,1\}$. (The special case where $C$=0 is handled analogously.)

It should be noted that the superindex $t$ is not stored in the memory of the automaton $A$, but it is not necessary: Automaton $A$ just stops checking when the input word $u^{2n}$ is completely read, and accepts only if it is able to reach the initial configuration of $M$.
\end{proof}

\section{Private alternation}
\label{sec:private-alternation}

Private alternation  \cite{Rei79} is a generalization of alternation \cite{CKS81} where the existential part does not see the universal part, and so more than one universal choice can follow the same existential choice, which allows  much better utilization of the existential choices, as will be seen soon. Originally, private alternation \cite{Rei79} was introduced as a two player game betweem universal and existential players, where the universal player is able to hide some information from the existential player. 

We will provide two different but equivalent definitions of realtime private alternating finite automata (PAFAs).  Firstly, we define the model as a single machine and a generalization of AFAs, and then, we give a definition adopted from the study of proof systems, where the existential player is a prover who is expected to provide the correct existential choices. Even though the first definition is fundamental, the second definition is more intuitive when presenting the algorithms as we do here. We believe that both definitions can be useful in   different contexts. 

Knowledgeable readers can see the analogy between alternation and Arthur-Merlin games \cite{Bab85}, and between private alternation and private-coin interactive proof systems (shortly IPS) \cite{GMR85,GMR89}. It may even be claimed that ``public alternation" would be a better name for what is now called ``alternation", to emphasize the main difference from private alternation. In her seminal PhD thesis, published as a book \cite{Co89}, Condon put all these models in a single framework. We refer the reader to her book for further discussions, and we construct our definition of the private alternating finite automaton (PAFA) model step by step. 

In  Condon's framework, the computation is governed privately by the universal player and the existential player is represented with only some finite number of states. Even though the existential player may also be given some private resources, Condon \cite{Co89} points out that we do not know whether this would bring any additional computational power. Therefore we also assume that the existential player is represented by some internal states. In \cite{Rei79, PR79}, the existential part can see the input head used by universal player, which restricts the computational power of the model to regular languages. Here we assume that, like in \cite{DSY15} and parallel to Condon's framework, there is only a single input head governed by the universal part, and the existential part does not see the head position.

A (realtime) PAFA $ P $ is a 8-tuple
\[
	P=( U,E,\Sigma,\delta_u,\delta_e,u_1,e_1,U_a),
\]
where
\begin{itemize}
	\item $ U $ is the set of universal states,
	\item $ E $ is the set of existential states,
	\item $ \delta_u $ and $ \delta_e $ are the transition functions for universal and existential parts (described later), respectively,
	\item $ u_1 \in U $ and $ e_1 \in E $ are the initial universal and existential states, respectively,
	\item $ U_a \subseteq U $ is the set of accepting (universal) state(s). 
\end{itemize}
Let $ w \in \Sigma^* $ be the given input. The automaton $ P $ reads the input as $ \cent w \dollar $. We can represent a configuration of  $ P $ as a triple $ (u,e,j) $, where $ u $ is the universal state, $ e $ is the existential state, and $ j $ is the position of head. The computation is governed by the universal part and it has the full knowledge of configuration. Moreover, the universal part, can update both the universal state and the existential state. Formally, the transition function is defined as 
\[
	\delta_u(u,e,\sigma) \rightarrow \mathcal{P}(U \times (\{\varepsilon\} \cup E)),
\]
where $ u \in U $, $ e \in E $, and $ \sigma \in \tilde{\Sigma} $. So, when the automaton is in universal state $ u $ and existential state $ e $ and reads symbol $ \sigma $, the universal part can make more than one transition and each transition can be either $ (u',\varepsilon) $ or $ (u',e') $, where $ u' \in U $ and $ e' \in E $. In the first type of transition, only the universal state is updated to $ u' $. The computation then continues with the next symbol in the universal part. There is no interaction with the existential part in this case. In the second type of transition, the existential state is also updated to $ e' $, furthermore,  an existential transition of the form 
\[
	\delta_e(e') \rightarrow \mathcal{P}(E)
\]
is implemented
before reading the next symbol.
Thus, the universal part can interact with the existential part in two steps. Then, the computation continues again with the universal part. The decision of ``acceptance'' is given when the universal state is a member of $ U_a $ at the end of the computation. Similar to AFAs, the whole computation can be represented as a single tree, where each node represents a configuration and each edge represents a transition from one configuration to another one. Any inner node is called universal (resp., existential) if it makes a universal (resp., existential) choice. A leaf is called accepting if its configuration contains an accepting state and it is called rejecting, otherwise. 

An AFA accepts an input if and only if there is an accepting subtree which is a subtree that (i) contains all children of a universal node and exactly one child of an existential node, and then, (ii) all leafs are accepting. The idea of keeping one child for each existential node can also be seen/called as an existential strategy. Therefore, we can also say that each existential strategy defines a subtree, i.e. we remove all the existential choices from the main computational tree not contained in the existential strategy. Then, we can say that, for an accepted input, there is always an existential strategy which defines an accepting subtree, or, by following this existential strategy, all possible universal choices lead us to the decision of ``acceptance''. 

The acceptance condition for PAFAs can be given similarly but an accepting subtree is obtained differently since the processing of an existential strategy is different. We refer to \cite{Co89} for more formal details. 

\begin{figure}[!ht]  
	\centering
	\begin{minipage}{0.9\textwidth}
	\fbox{
  \centerline {
  	\ifpdf
		\includegraphics[width=0.8\textwidth]{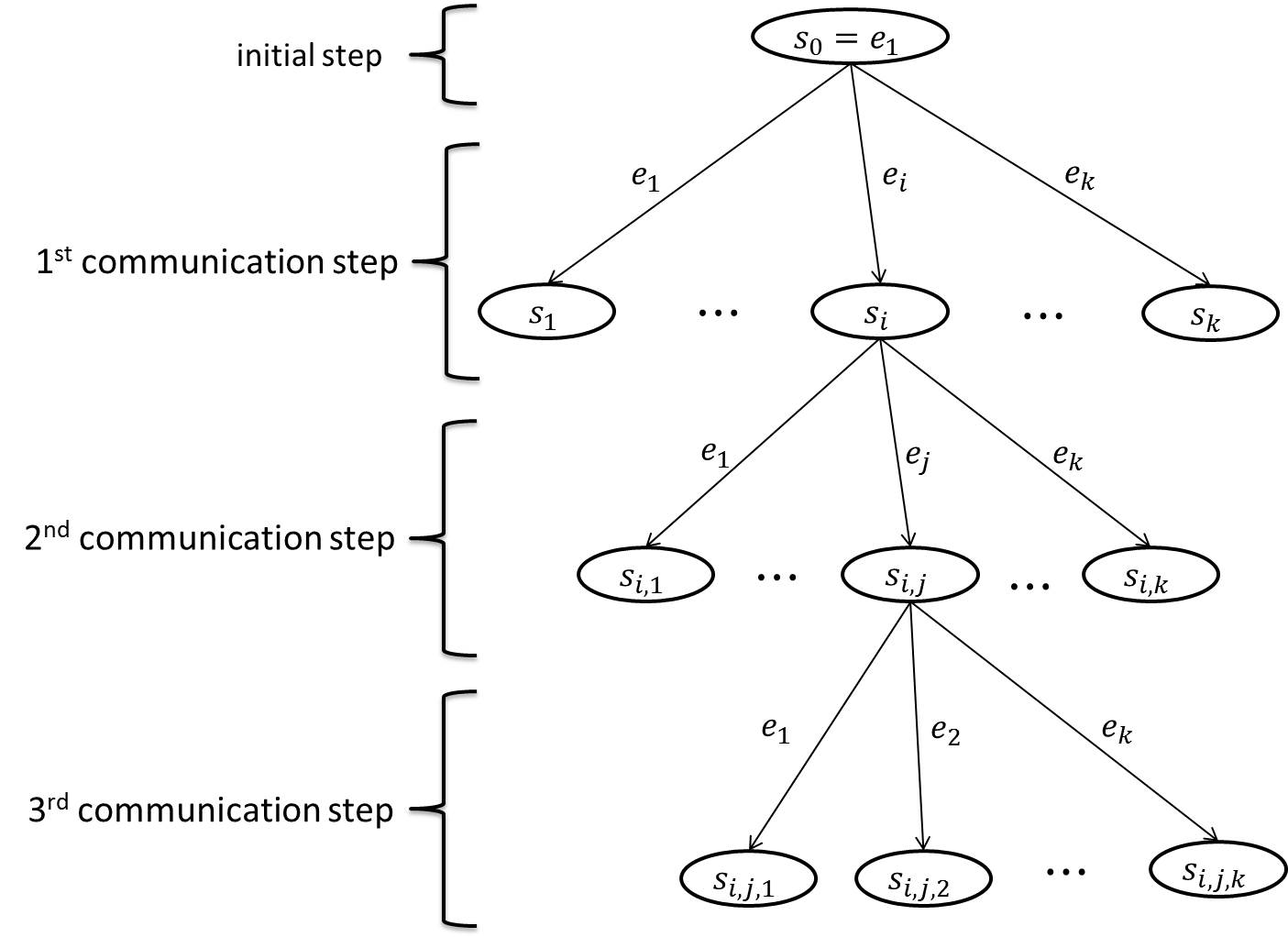}
	\else
  		\includegraphics[width=0.8\textwidth]{es}
	\fi} 
  }
  \end{minipage}
  \caption{Three communication steps}
\label{fig:communication-steps}
\end{figure}

Remember that the existential part of $ P $ can see only the transitions on existential states, which is set to $ e_1 $ at the beginning of the computation. During the computation, the universal part updates an existential state and this triggers that existential part that makes its own updates. Then, a single existential strategy can be detailed below. The reader can follow the explanations from Figure \ref{fig:communication-steps}.

We will use variables $ s $s to show the selected existential states. Let $ k $ be the size of $ E $. The initial state $ s_0 = e_1 $. Now we enumerate the steps when there is an interaction between universal and existential parts separately, i.e., more precisely, when the automaton switches from an universal state to an existential state. We call them as communication steps. 

In the first communication step, the universal part can have $ k $ different choices when updating the existential part, i.e. $ \{ e_1,\ldots,e_k \} $. For each of them, the existential part makes its own updates. Even though there are more than one update, we fix one of them for a single strategy. We denote the new existential states as $ s_1,\ldots,s_k $, where $ s_i $ is the corresponding update of the existential part when universal part selects $ e_i $ and $ 1 \leq i \leq k $:
\[
	s_0=e_1 \xrightarrow{universal~part} e_i \xrightarrow{existential~part} s_i \in \{e1,\ldots,e_k\}.
\]
Remark that there can be two different indices $ 1\leq i \neq j \leq k $ such that $ s_i = s_j $. Moreover, this first communication step can correspond to more than one computational step in the ``main'' tree, representing the whole computation. (The implementation of each node at the same level in Figure~\ref{fig:communication-steps} can be in different steps.) For example, the universal part can change the existential state to $ e_i $ for the first time in computational steps 3, 7, and 11. Then, there are three existential inner nodes in the main computational tree with different levels. \textit{When obtaining a subtree, with respect to our existential strategy, only the existential choices $ s_i $ are kept outgoing from those inner nodes, even though they can be at the different levels in the main tree.} In this way, the different universal choices can follow the same existential strategy at different computational steps. In other words, when accessing the different part of the inputs, we can still follow the same existential choices. This is the key idea behind  private alternation. 

In the second communication step, we repeat the same procedure. So, if our existential state is $ s_i $, then the new existential states will be $ s_{i,1},s_{i,2},\ldots,s_{i,k} $. In the third communication step, if the existential state is $ s_{i,j} $, then the new existential states will be $ s_{i,j,1},s_{i,j,2},\ldots,s_{i,j,k} $, where $ 1 \leq i,j, \leq k $. We continue in the same way for other communication steps. 

Each existential strategy for $ P $ defines a subtree such that all universal choices are kept and, when pruning the existential inner nodes, the only remaining existential choice must be the same for the nodes that (i) corresponds to the same communication step and (ii) contains the same existential state. An input is accepted by $ P $ if and only if there exists an existential strategy that defines an accepting subtree (whose leaves are accepting ones). 

\subsection*{Example:} We continue with a simple example. Let $ a^mb^n $ be the promised input, where $ m,n > 0 $, and, $ P $ be a PAFA that separates the case $ m=n $ from $ m \neq n $. We have 4 universal and 2 existential states, $ \{ u_1,u_2,u_r,u_a \} $ and $ \{e_1,e_2\} $, respectively, where $ u_1 $ and $e_1$ are initial states and $ u_a $ is the only accepting state. 

The main idea is that the universal part expects from the existential part to nondeterministically pick a length, say  $ l $, which can be used to compare the equality of $ m $ and $ n $ in the universal branches. Remark that if $ m=n $, then $ l=m=n $ will be an evidence for the equality. However, if $ m \neq n $, then $ l $ cannot be equal to both of $ m $ and $ n $ and so the universal part cannot be convinced for $ m=n $.  

After reading $ \cent $, $ P $ universally splits into two paths, with the following transition:
\[
	\delta_u(u_1,e_1,\cent) \rightarrow \{ (u_1,\varepsilon),(u_2,\varepsilon) \}.
\]
The first (resp., second) path reads $a$s (resp., $b$s) and then gives a decision. The universal part never branches again and follows only deterministic transitions. The details are given below.

The first path: For each $a$, $ P $ switches from $ (u_1,e_1) $ to $ (u_1,e_1) $ and the existential part switches from $ e_1 $ to $ e_1 $ and $ e_2 $:
\begin{eqnarray*}
	\delta_u(u_1,e_1,a) & \rightarrow & \{ (u_1,e_1) \}
	\\
	\delta_e(e_1) & \rightarrow & \{ e_1,e_2 \}.
\end{eqnarray*}
Here switching $ e_2 $ means that the existential part finishes its task, i.e. $l$ is picked. Thus, if $ P $ is in $ (u_1,e_2) $ and reads an $ a $, i.e. $ l \neq m $, then it switches to $ (u_r,\varepsilon) $:
\[
	\delta_u( u_1,e_2,a ) \rightarrow \{ ( u_r,\varepsilon ) \}.
\]
Once $ P $ enters $ u_r $, the decision of ``rejection'' is given by staying in $ u_r $ until the end of the input. If $ P $ reads the first $ b $, then the first path ends with the following transitions:
\begin{eqnarray*}
	\delta_u(u_1,e_1,b) & \rightarrow & \{ (u_r,\varepsilon) \}
	\\
	\delta_u(u_1,e_2,b) & \rightarrow & \{ (u_a,\varepsilon) \}.
\end{eqnarray*}
The first transition means $ l>m $ and the second transition means $ l=m $. Thus, in the former case, the decision of ``rejection'' is given as described above. In the latter case, $ P $ enters $ u_a $, and then the decision of ``acceptance'' is given by staying in $ u_a $ until the end of the input. So, the first path ends with the decision of ``acceptance'' only if there is an existential strategy that chooses $ e_1 $ $ (m-1) $ times and then a single $ e_2 $, i.e. $ l=(m-1)+1 = m $.

The second path: When in $ u_2 $, $ P $ does not do anything while reading $ a $s. When starting to read $ b $s, the same strategy given for the first path is applied:
\begin{eqnarray*}
	\delta_u(u_2,e_1,b) & \rightarrow & \{ (u_2,e_1) \}
	\\
	\delta_u(u_2,e_2,b) & \rightarrow & \{ (u_r,\varepsilon) \}.
\end{eqnarray*}
After reading $ \dollar $, it is expected to see $ e_2 $ for the decision of ``acceptance''. 
\begin{eqnarray*}
	\delta_u(u_2,e_1,\dollar) & \rightarrow & \{ (u_r,\varepsilon) \}
	\\
	\delta_u(u_2,e_2,\dollar) & \rightarrow & \{ (u_a,\varepsilon) \}.
\end{eqnarray*}
So, the second path ends with the decision of ``acceptance'' only if there is an existential strategy that chooses $ e_1 $ $ (n-1) $ times and then a single $ e_2 $, i.e. $ l=(n-1)+1 = n $.

It is clear that, for the case of $ m=n $, there exists an existential strategy (picking $ l=m=n $) such that both paths give the decision of ``acceptance''. But, for the case of $ m \neq n $, there is no existential strategy that leads each path to with the decision of ``acceptance''. For example, the existential strategy must pick $ (m-1) $ $e_1$s and then a $ e_2 $ to lead the first path to giving the decision of ``acceptance''. But, then, the second path gives the decision of ``rejection''.

In this example, the universal part does not switch to any existential state except the initial one. In fact, this is a standard restriction for private alternation, and such restricted a PAFA is called a blind alternating finite automaton (BAFA). The universal part of a BAFA always switches to a fixed existential state and so the interaction is one-way.

With this example, we can also conclude that BAFAs are more powerful than AFAs.

\begin{lem}
	\label{lem:anbn}
	BAFAs can recognize the language $ \{a^nb^n  \mid n>0 \} $.
\end{lem}
\begin{proof}
	Whether a given input is of the form $ a^+b^+ $ can be deterministically checked, and so, the result follows from the above example.
\end{proof}

\begin{cor}
	\label{cor:more}
	BAFAs can recognize all regular and some nonregular languages.
\end{cor}

As also seen from the above example, any existential strategy of a BAFA can be seen as a sequence of existential states. Whenever the universal part communicates with the existential part, it reads a state from this sequence. When considering analogy between private alternation and IPS, blind alternation corresponds to one-way IPS. In one-way IPS, the communication with a prover can also be represented as a single certificate (then the prover is omitted), which is written on a separate one-way read-only tape (e.g. see \cite{SY12C,SY14A}). Similarly, for BAFAs, a sequence of existential states (representing an existential strategy) can be considered as a certificate which is written to a separate tape one-way accessible by the automaton. 

Now, we give the alternative definitions of BAFAs and PAFAs. A BAFA is an automaton having  read-only one-way infinite input and certificate tapes. The head of the input tape is realtime and the head of the certificate tape is one-way. Formally, a BAFA $B$ is a 6-tuple
\[
	B = ( U, \Sigma, \Gamma,\delta,u_1,U_a ),
\]
where
\begin{itemize}
	\item $U$ is the set of (universal) states,
	\item $ \Gamma $ is the finite certificate alphabet, (with $\# \notin \Gamma $ as the blank symbol),
	\item $ \delta $ is the transition function (described later),
	\item $ u_1 \in U $ is the initial state, and,
	\item $ U_a \subseteq U $ is the set of accepting state(s). 
\end{itemize}
Let $ w \in \Sigma^* $ be the given input and $ c \in \Gamma^* $ be the given certificate. The input is written on the input tape as
\[
	\cent w_1 w_2 \cdots w_{|w|} \dollar.
\]
In fact, the remaining tape squares contain the blank symbol but it is not essential since the computation is terminated after reading $\dollar$. The certificate is written on the certificate tape as
\[
	c_1 c_2 \cdots c_{|c|} \# \# \cdots
\]
Remark that, for our realtime models, we assume that each certificate is a finite string. For the models that do not have a time bound, the certificate can also be infinite since the communication does not need to be finite if the running time can be infinite. The transition function is defined as follows:
\[
	\delta (u,\sigma,\gamma) \rightarrow \mathcal{P}(U \times \{ 0,1 \}),
\]
where $ u \in U $, $ \sigma \in \tilde{\Sigma} $, and $ \gamma \in \tilde{\Gamma} $. When the automaton is in state $u$ and reads $ \sigma $ and $ \gamma $ on the input and certificate tapes, respectively, it makes more than one transition in the form $ (u',d) $ where $ u' \in U $ is the new state and the position of certificate head is updated with respect to $ d \in \{0,1\} $, i.e. the head does not move if $ d = 0 $ and moves one square to the right if $ d = 1 $. Note that after each transition the input head moves one square to the right. The decision of ``acceptance'' is given if the automaton enters an accepting state after reading the right end-marker and the decision of ``rejection'' is given, otherwise. If each branch ends with the decision of the ``acceptance'', then the input is accepted by the automaton. If at least one branch ends with the decision of ``rejection'', then the input is rejected. Remark that, on the same input, we can get different decisions for different certificates.

A language $ L $ is said to be recognized by $ B $ if:
\begin{enumerate}
\item For a member $ w \in L $, there exists a certificate $ c_w \in \Gamma^* $ such that $ B $ accepts $ w $, and,
\item for a non-member $ w \notin L $, $ B $ always rejects $ w $ for any certificate $ c \in \Gamma^* $ is.
\end{enumerate}

When considering our example above, it is clear that the certificate $ a^{n-1}b $ is sufficient for the automaton to accept the input of the form $ a^nb^n $. On the other hand, none of the certificate $ a^{l-1}b $ leads us to accept the inputs $ a^nb^m $ ($ n \neq m $). 

In the alternative definition of PAFAs, we can use a tree certificate instead of a linear one. That is, a PAFA can have a tree one-way infinite tape where the head never goes  up, and each node of the tree contains a symbol from certificate alphabet. The certificate depth is assumed to be finite. When the certificate is finished, only the blank symbols are written on the tape. We also assume that the tree is $k$-ary for some $ k >1 $.  Then, the new transition function is as follows:
\[
	\delta (u,\sigma,\gamma) \rightarrow \mathcal{P}(U,\{ 0,1,\ldots,k \}),
\]
where the parameter $ d $ (see above) can take a value from the set $ \{ 0,1,\ldots,k \} $ and the head does not move if $ d=0 $ and the head picks the $ i $th child if $ d=i $ (we can assume that the children are labelled from left to right by $ 1, \ldots, k $). The acceptance condition is the same as BAFAs.

An IPS is said to  have \textit{perfect completeness} if the accepting probability for each member of the language is 1. If such an IPS has a time bound, then we can say that every probabilistic path ends with the decision of ``acceptance'' for inputs that are members of the language. It is clear that if we replace probabilistic choices with universal ones, then we can have a PAFA for the same language, i.e. each path ends with the decision of ``acceptance'' for the members by help of some certificates, and, at least one path ends with the decision of ``rejection'' for non-members whatever the certificate is. The same connection is not clear for space-bounded IPSs since the accepting probability 1 can be obtained by some infinite loops. We present this connection as a lemma in a more general form.

\begin{lem}
	\label{lem:perfect}
	Let $ M $ be a time-bounded probabilistic verifier that can access some (linear or tree) certificates as described above, and $ L \subseteq \Sigma^* $ be a language defined by $ M $ as follows:
	\begin{enumerate}
		\item each $ w \in L $ is accepted with probability 1 by using a certificate, and,
		\item each $ w \notin L $ is accepted with probability less than 1 for any  certificate.
	\end{enumerate}
	Then, $ L $ is recognized by a private alternating machine $ M' $, which is simply obtained from $ M $ by changing probabilistic states with universal ones. If the certificates are restricted to be linear, then $ M' $ is a blind alternating machine. 
\end{lem}

\begin{cor}
	\label{cor:twin}
	BAFAs can recognize the language  $ \mathtt{TWIN} = \{ w2w \mid w \in \{0,1\}^* \} $.
\end{cor}
\begin{proof}
	In \cite{SY14A}, a realtime finite state probabilistic verifier using linear certificate is given for this language such that each member is accepted with probability 1 and each non-member is accepted with probability at most $ \frac{2}{3} $. Due to Lemma \ref{lem:perfect}, we obtain the result.
	
	We also provide an explicit proof. Let $ w= w_1 2 w_2 $ be the given input where $ w_1,w_2 \in \{0,1\}^* $. Any input not in this form can be deterministically rejected. 
	
	Let $ c \in \{0,1\}^* $ be the certificate. At the beginning of the computation, the automaton universally splits into two paths. In the first (resp., second) path, it checks whether $ w_1 = c $ (resp., $w_2=c$). Thus, if $ w \in \twin $, then by help of the certificate $ c=w_1=w_2 $, the computation ends with an accepting state in both paths. If $ w \notin \twin $, then the computation ends with a rejecting state in at least one of the paths, since $ c $ cannot be equal to $ w_1 $ and $ w_2 $ at the same time when $ w_1 \neq w_2 $.
\end{proof}

\newcommand{\upow}{\mathtt{UPOWER}}

We now show that BAFAs can also recognize nonregular unary languages. In the previous version of this paper \cite{DBY14A}, we showed that $ \upow = \{1^m \mid m=2^n \text{ for some }n \geq 0 \}$ is recognized by a PAFA. Here we present a simpler proof. 

\begin{thm}
	\label{thm:upower}
	There is a BAFA $ B $ recognizing $ \upow $.
\end{thm}
\begin{proof}
	Let $ a^m $ be the input with length at least 4. (The shorter strings are checked deterministically.) The certificate is of the form  $ (a^+b^+)^+a^* $, being composed of blocks of $ a $s and $ b $s. Let $ k $ be the number of blocks and let $ t_1,\ldots,t_k $ represent the number of symbols in each block, respectively. For example, for certificate $ a^8b^4a^2b $, there are four blocks (k=4) and $ t_1=8 $, $ t_2=4 $, $ t_3=2 $, and $ t_4=1 $.
	
	The automaton $ B $ branches universally to  make $ k+1 $ tests:
	\begin{itemize}		
		\item $ test_1 $ checks whether $ m= (t_1) + t_1 $,
		\item $ test_i $ checks whether $ m = (t_1+\cdots+t_i)+t_i $ for $ i \in {2,\ldots,k} $, and,
		\item $ test_{k+1} $ checks whether $ m=1+\sum_{j=1}^k t_j $,
	\end{itemize}
	
	It is easy to see that if all tests are successful, then $ m=2 t_1 $, $ t_1=2t_2 $, \dots, $ t_{k-1} = 2 t_k $, and, due to the last test,  
	\[
		m = 1 + \frac{m}{2} + \frac{m}{4} + \cdots + \frac{m}{2^k}.
	\]
	From this equation, we get $ m=2^k $ after straightforward calculations. 
	
	The automaton $ B $ makes the last test by reading the input and certificate strings with the same speed. If the certificate string is finished before the input string and there remains exactly one  input symbol, then the test is passed and the decision of ``acceptance'' is given. Otherwise, the test is failed, and the decision of ``rejection'' is  given in this universal path. We call this universal path as the main path.
	
	When the $i$-th block starts, $ B $ creates a new universal path from the main path to make $ test_i $ ($ 1 \leq i \leq k $).   
	For each symbol in the $i$-th block, $ B $ tries to read two input symbols in this new path. If the whole input and the $ i $-th block is finished at the same time, then the test is passed and the decision of ``acceptance'' is given in this universal path. Otherwise, the test is failed and the decision of ``rejection'' is given.
	
	Remark that each $test_i$ ($ 1 \leq i \leq k $) is implemented with the same set of states. Therefore, even though there are $ k+1 $ different tests, the number of states are still finite. 
	
	If $ m=2^n $ for some $ n>1 $, then all test will be passed successfully by the help of a certificate with $ n $ blocks such that
	\[
		t_1 = 2^{n-1}, t_2=2^{n-2}, \ldots, t_n=2^{n-n}=1.
	\]
	
	If $ m $ is not a power of 2, then one of the tests will fail for any certificate. The reader can try to construct a certificate by passing each test in the given order and she will see that the construction will fail at one of the steps.
\end{proof} 

In the previous version of this paper, we also presented a realtime private alternating counter automaton for language $ \usqr  = \{1^m \mid m=n^2 \text{ for some } n \geq 0 \}$. Now we show that  BAFAs can recognize this language without using any counter. Remark that $\usqr$ seems harder than $ \upow $. For example, it was left open whether two-way deterministic counter automata can recognize $ \usqr $ by Petersen \cite{Pet94}. In fact, the only type of counter automaton known to recognize it until now is an exponential-time quantum version with two-way access to the input \cite{Yak13B}.

\begin{thm}
	\label{thm:usqr}
	There is a BAFA $ B $ recognizing $ \usqr $.
\end{thm}
\begin{proof}
	The proof is similar. This time we use the fact that the sum of the first $ n $ odd numbers is $ n^2 $, i.e. $ 1+3+5+\cdots+2n-1=n^2 $. For this purpose, we use consecutive natural numbers on the certificate. The reader may question why we do not directly use consecutive odd numbers on the certificate. Our testing procedures involve the head pausing for one or two steps while traversing  portions of the certificate tape, and so  the certificate would be correspondingly shorter than the input.
	
	Let $ a^m $ be the input with length at least 4. (The shorter strings are checked deterministically.) The certificate is composed of blocks of $ a $s and $ b $s as described in the  proof of Theorem \ref{thm:upower}. Let $ k $ be the number of blocks and let $ t_1,\ldots,t_k $ represent the number of symbols in each block, respectively.
	
	The automaton $ B $ branches universally for all values of $i$ mentioned below to make the following tests:
	\begin{itemize}
		\item $ test_1 $ checks whether $ t_1 = 1 $,
		\item $ test_i $ checks whether $ t_i=t_{i-1}+1 $ for $ i \in \{2,\ldots,k\} $,
		\item $ test_{k+1} $ checks whether $ m=1+\sum_{j=1}^k (2t_j+1) $.  
	\end{itemize}
	It is easy to see that if all tests are passed successfully, then $ t_1=1, t_2=2, \ldots, t_k=k $, and, due to the last test,
	\[
		m= 1+3+\cdots+2k+1 = (k+1)^2.
	\]
	
	The first test is trivial. For $ i \in \{2,\ldots,k\} $, $ test_i $ is implemented in three phases by reading symbols parallelly from the input and certificate tapes in the following carefully synchronized manner: 
	\begin{itemize}
		\item Phase 1: $ B $ starts with reading a single input symbol. Then for all  $ j < i-1 $, it attempts to read  $ 2t_j+1 $ input symbols when consuming  the $ j $'th certificate block, rejecting if the input ends prematurely.
		\item Phase 2: $ B $ splits universally into two paths when starting to consume the ($i-1$)'st certificate block. In the first path, $ B $ tries to read $ 3t_{i-1}+3+t_i $ input symbols while consuming the $ (i-1) $'st and $ i $'th blocks. In the second path, $ B $ tries to read $ t_{i-1}+1+3t_i $ input symbols while consuming the same blocks.
		\item Phase 3:  While processing the remainder of the input, $ B $ tries to read $ 2t_j+1 $ input symbols as it consumes the $j$'th certificate block (for all $ j > i) $ in both of the computational branches created in Phase 2. 
	\end{itemize}
	If the certificate and the input are finished at the same time, then the test is passed and the decision of ``acceptance'' is given. Otherwise, the test is failed and the decision of ``rejection'' is given.  
	
	During $ test_i $, the operations performed by the two branches created in Phase 2 differ only when reading the $ (i-1) $'st and $i$'th blocks of the certificate. If both paths finish the computation at the same time, then the following equality is satisfied:
	\[
		3t_{i-1}+3+t_i = t_{i-1}+1+3t_i.
	\]
	It follows that $ t_{i-1}+1=t_i $. When $ test_i $ is passed, $ B $ reads 
	\[
		\sum_{j=1}^k 2t_j+1
	\]
	  input symbols in both paths, since the input symbols read for the $ (i-1) $'st and $i$'th blocks of the certificate is $ t_{i-1}+1+3t_i = (2t_{i-1}+1)+(2t_i+1) $. Thus, each $ test_i $ also implements $ test_{k+1} $, leaving no need to implement $ test_{k+1} $ separately.
	  
	If $ m=n^2 $ for some $ n>1 $, then all tests will be passed successfully by help of the certificate with $ n-1 $ blocks such that
	\[
		t_1 = 1, t_2=2,\ldots,t_{n-1}=n-1.
	\]
	
	If $ m $ is not a perfect square, then one of the tests will fail for any certificate.
\end{proof}

\begin{thm}
	\label{thm:PAFA-emptiness}
	The emptiness problem for BAFAs is undecidable.
\end{thm}
\begin{proof}
Assume that there exists a decision procedure for this problem. We can then decide whether a given TM $  M $  accepts an input $ x $ as follows: Construct a BAFA $ B $, described below, which accepts a string only if it is an accepting computation history of $ M $ on $ x $, (i.e. a string composed of segments representing the configurations that $ M $ would go through when started on $ w $, all the way to an accepting configuration) properly encoded using 0's and 1's, use the emptiness test on $ A $ to see whether $ M $ accepts $ w $, and announce the opposite of what the emptiness test says.

Let $ C_0,C_1,\ldots,C_n \in \{0,1\}^* $ be the  configurations that  $ M $ would go through when started on $ x $ such that $n$ is the total number of steps, $ C_0 $ is the initial configuration, $ C_n $ is the final and accepting configuration, and $ C_i $ is the configuration after the $ i $-th step, where $ i \in \{1,\ldots,n\} $. Of course, such a finite sequence exists if and only if  $ M $ accepts $ x $.

$ B $ accepts only the string
\[
	x' = \kappa \kappa C_0 \kappa \kappa C_1 \kappa \kappa C_2 \kappa \kappa \cdots  \kappa \kappa C_n \kappa \kappa
\]
by using the  certificate
\[
	c' = C_0 \kappa C_1 \kappa \cdots \kappa C_n,
\]
where $\kappa$ is a separator symbol.
Even though the accepted string and its certificate describe the computation history of $ M $ on $ x $, we use these  separators in the accepted string for possible delays on the certificate tape.

The strategy of $B$ is as follows. Let $ w = \kappa\kappa w_0 \kappa\kappa w_1 \kappa \kappa \cdots \kappa \kappa w_m \kappa\kappa $ be the given input where $ m>0 $, and $ w_i \in\{0,1\}^* $ for $ i \in \{0,\ldots,m\} $. If the input is not in this form, then it is rejected deterministically. Let $ c = c_0 \kappa c_1 \kappa \cdots \kappa c_l $ be the certificate, where $ l>0 $, and  $c_i \in\{0,1\}^* $ for $ i \in \{0,\ldots,l\} $. If the certificate is not in this form, then the input is rejected deterministically.

$ B $  splits universally to two paths at the beginning of the computation. In the first path, $ B $ deterministically checks whether $ w_0=c_0=C_0 $ at the beginning of the computation with the help of internal states encoding $C_0 $,   and reading $ w_0 $ and $ c_0 $ in parallel. If this comparison fails, the decision of ``rejection'' is given. Otherwise,   $ B $ tries to check one by one whether each pair $ (w_i,c_i) $ is identical or not, for each $ i>0 $. If one of these comparisons fails, then the decision of ``rejection'' is given. If $ l \neq m $, then the reading of one tape  finishes earlier than the other,  at which point the decision of ``rejection'' is given. If all comparisons are successful and $ l=m $, then the decision of ``acceptance'' is given in this path.  

In the second path, $ B $ checks whether $ w_1 $ is a description of the legal successor of $ c_0 $ according to the transition function of $M$.  If not, the decision of ``rejection'' is given. Such a check can be performed by encoding the allowed ``windows'' of substring changes that can occur in the configuration of $M$ in a single step in the internal states of $B$, and reading the input and certificate tapes in parallel. As long as none of comparisons fails and both the input and the certificate  strings are not finished, this path tries to repeat the same test for the pairs
\[
	(w_2,c_1), (w_3,c_2), \ldots, (w_{i+1},c_i), \ldots .
\]
If the last pair is $ (w_{m},c_{l-1}) $, the check is successful, and  $ w_m $ is an accepting configuration, then the decision of ``acceptance'' is given. In any other case, the decision of ``rejection'' is given.

It is clear that if $ w=x' $ and $ c=c' $, then $ B $ accepts. If $ w \neq x' $, then there is at least one inconsistency on $ w $ and $ B $ can easily detect this in one of the paths: if $ w $ and $ c $ do not contain an identical sequence of  configurations, then this is detected in the first path. If such a $ w $ and $ c $ contain the same configurations, then either at least one of the configurations is not  the legal successor of the previous configuration, or the very last configuration is not accepting.
\end{proof}

\section{Quantum alternation}

We refer the reader to \cite{NC00,SayY14,AY15A} for the basics of quantum computation and quantum automata. A quantum finite automaton (QFA) \cite{Hir10,YS11A} is a quintuple
\[	
	M = (Q,\Sigma,\{ \mathcal{E}_{\sigma} \mid \sigma \in \tilde{\Sigma} \},\ket{v_0},P), 
\]
where $Q=\{q_1,\ldots,q_n\}$ is the set of states for some $ n>0 $, $\mathcal{E}_{\sigma} = \{ E_{\sigma,1},\ldots,E_{\sigma,l_\sigma} \} $ is the superoperator for the symbol $ \sigma \in \tilde{\Sigma} $ composed by $l_\sigma$ operation elements, $ \ket{v_0}\in \{\ket{q_1},\ldots,\ket{q_n}\} $ is the initial state, and $P = \{ P_a,P_r \}$ is the measurement operator applied after reading the right end-marker. An input is accepted if the outcome ``a'' of $P$ is observed.
For any given input $w\in \Sigma^*$, the computation of $ M$ can be traced by a $ |Q| \times |Q| $-dimensional density operator (mixed state)\footnote{A pure state is a basis state or a linear combination of basis states with norm 1. A mixed state is a mixture of pure states such that the system is in each pure state with a nonzero probability and the summation of all probabilities is 1, i.e. $ \{ (p_i,\ket{\psi_i} \mid 1 \leq i \leq t \mbox{ and } \sum_{i=1}^t p_i =1 ) \} $ for $t>0$ pure states.}:
\[
	\rho_j = \mathcal{E}_{w_i} (\rho_{j-1}) = \sum_{k=1}^l E_k \rho_{j-1} E_k^{\dagger} ,
\] 
where $\rho_0 = \ket{q_0} \bra{q_0} $ and $ 1 \leq j \leq |w| $, and the acceptance probability of $ M$ on $w$ is 
\[
	f_{ M}(w) = Tr( P_a \rho_{|w|} ).
\]

The nondeterministic QFA (NQFA) model can be defined using an acceptance mode known as \textit{positive one-sided unbounded error} \cite{ADH97,YS10A}: A language $L$ is said to be recognized by a NQFA $ N $, where $N$ is a QFA, if each member of $L$ is accepted by $ N $ with a positive probability, and each member of $\overline{L}$ is accepted by $ N $ with zero probability.

\begin{thm}
	The emptiness problem for NQFAs is decidable, where the automata are defined using algebraic numbers as transition amplitudes in the operation elements. Moreover, if the transition amplitudes are restricted to rational numbers, we can also give a time bound.
\end{thm}
\begin{proof}
	We can easily design a 1-state QFA named, say, $ M $, with only rational amplitudes,  which accepts all inputs with probability 0. Let $ N $ be the NQFA which we are supposed to test. Firstly, we assume that $ N $ is defined with only rational numbers. The equivalence problem for QFAs, i.e. deciding whether  two given QFAs on the same input alphabet have identical acceptance probability functions, is solvable in polynomial time if the automata are defined with rational amplitudes \cite{Tze92,LQZLWM12}. Therefore, we can easily design a polynomial-time algorithm that takes $ M $ and $ N $ as  input and  determines whether they are equivalent or not. They are equivalent if and only if $ N $ recognizes the empty set. 
	
	We now allow $ N $'s operation elements to contain arbitrary algebraic numbers. The minimization problem for QFAs, i.e. taking a QFA as the input and then outputting a minimal equivalent QFA, is also solvable if the amplitudes are restricted to algebraic numbers \cite{MQL12}. Therefore, we can design an algorithm that takes $ N $ as the input, constructs its equivalent minimal QFA, and accepts only if that minimal machine has a single state, like the obviously minimal QFA $M$ mentioned above. 
\end{proof}

The universal QFA (UQFA) model, which can be considered as the ``complement'' of the NQFA, is defined using an acceptance mode known as \textit{negative one-sided unbounded error} \cite{YS10A}. A language $L$ is said to be recognized by a UQFA $ U $ if each member of $L$ is accepted by $ U $ with probability 1, and each member of $\overline{L}$ is accepted by $ N $ with probability less than 1. For a given QFA with rational amplitudes, say $ M$, and $ p \in [0,1] $, the problem of whether there is a string accepted by $ M$ with probability $ p $ is undecidable. (Note that this result can be obtained even for the simplest known  QFA model \cite{MC00}, known as measure-once or Moore-Crutchfield QFA \cite{DJK05,BJKP05}.) Therefore, the emptiness problem for UQFAs is undecidable.

\begin{cor}
	The emptiness problem for UQFAs with rational amplitudes is undecidable.
\end{cor}

The class of languages recognized by NQFAs (resp., UQFAs) form a superset of the regular languages \cite{YS10A}, called exclusive (resp., co-exclusive) stochastic languages ($\sf S^{\neq}$ (resp., $\sf S^{=}$)) \cite{Paz71}.  $ \sf S^{\neq} $ and $\sf S^{=}$ do not contain any nonregular \textit{unary} languages \cite{SS78}. Therefore, it is interesting to ask whether alternation between the capabilities of the two could lead us to recognizing some nonregular unary languages. It is known that such ``two-alternation" is sufficient for recognizing the well-known $\sf NP$-complete language $\tt SUBSETSUM$ (or $\tt KNAPSACK$) \cite{Yak13A,Yak16A}. It is therefore reasonable to  expect that two-alternation is useful for some unary nonregular languages. 
In the following, we will first describe a model of quantum alternation that embodies these ideas, and then present a result for $ \tt \mathtt{USQUARE} $, using the methods introduced in \cite{Yak13A,Yak16A}.

\begin{figure}[!ht]  
	\centering
	\begin{minipage}{0.9\textwidth}
	\fbox{
  \centerline {
  	\ifpdf
		\includegraphics[width=0.75\textwidth]{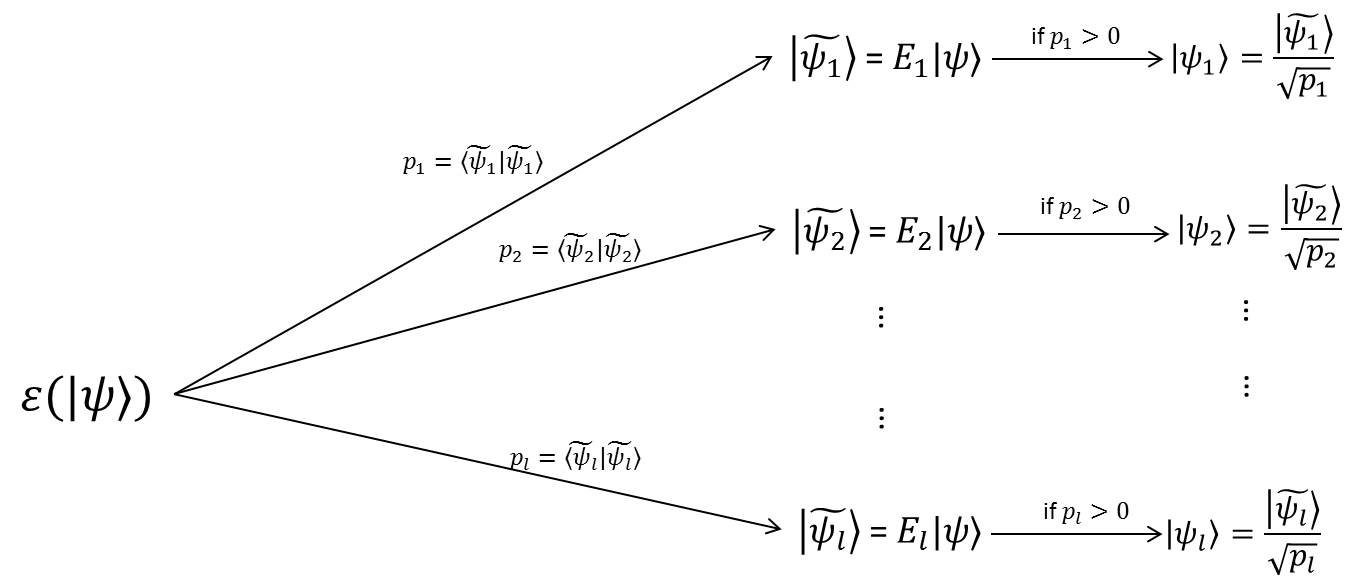}
	\else
  		\includegraphics[width=0.75\textwidth]{superoperator-from-jpeg}
	\fi}
  }
  \end{minipage}
  \caption{The pure states after applying a superoperator with $l$ operation elements}
\label{fig:superoperator-purestates}
\end{figure}

An alternating quantum finite automaton (AQFA) has both classical and quantum states \cite{Yak13A}. The classical states are once again either existential or universal, with some of them designated as accepting states. The computation is controlled by the classical states. Each step has two phases. In the first phase (quantum phase), depending on the current symbol and classical state, a superoperator is applied to the quantum state, and an outcome having non-zero probability is observed. Figure \ref{fig:superoperator-purestates} represents the outcomes of a superoperator applied to a pure quantum state, where the indices of the operators denote the outcomes. Each outcome with non-zero probability becomes a branch of the computation.  
In the second phase (classical phase), the new classical state is determined by the current symbol, the classical state, and the outcome obtained in the first phase. Similar to the previous models, we assume that the automaton spends two steps on each symbol. The computation of an AQFA can be shown as a tree such that each node represents a pair consisting of a classical state and a pure quantum state, and the edges represent the transitions with non-zero probabilities. Note that the exact magnitudes of these probabilities are not significant. We associate each inner node with either a ``$\vee$'' or a ``$\wedge$'', depending on whether its corresponding classical state is existential or universal, respectively. When evaluating the tree, each leaf in which the input is accepted (rejected) takes the value of ``true'' (``false''). These halting decisions are given by checking whether the computation ends in a classical accepting state. The value of root can be evaluated from bottom to top: Any existential inner node (assigned a ``$\vee$'') takes the value of ``true'' if at least one of its children has taken the value of ``true'', and any universal inner node (assigned an ``$\wedge$'') takes the value of ``true'' if all of its children have taken the value of ``true''. The input is accepted if and only if the root takes the value of ``true''.  Note that classical alternation is a special case of quantum alternation, and NQFAs and UQFAs are just AQFAs with only existential and universal states, respectively.

\begin{thm}
\label{thm:AQFA-usquare}
	The language $\tt \mathtt{USQUARE}$ can be recognized by a AQFA, say $A$, with two alternations.
\end{thm}
\begin{proof}	
	We use the programming techniques given in \cite{Yak16A}. The automaton has two classical existential states $ e $ and  $ e_r $ and two classical universal states $ u $ and $ u_a $.  The initial classical state is $ e $ and the only accepting state is $ u_a $. States $ e $ and $ u $ are the main existential and universal states, respectively, and, states $ e_r $ and $u_a$  are auxiliary.
    
    $ A $ switches from state $ e $ to states $ e $, $ e_r $, and $ u $. The transitions to $ e_r $ can be ignored since once $ e_r $ is entered, $ A $ stays in $ e_r $ until the end of the computation (and the decision of ``rejection'' is given at the end).
    
    $ A $ switches from state $ u $ to either to states $ u $ and $ u_a $ or only to state $ u_a $. Once $ u_a $ is entered, $ A $ stays in $ u_a $ until the end of the computation (and so the decision of ``acceptance'' is given). Therefore, the input  can be accepted after  a sequence of states of the form
    \[
    	\begin{array}{llllllllllll}
    		e \rightarrow \cdots \rightarrow e \rightarrow & u & \rightarrow & u & \rightarrow  & \cdots & \rightarrow   & u & \rightarrow  & u & \rightarrow & u_a \\
            & \downarrow & & \downarrow & & & & \downarrow \\
            & u_a & & u_a & & & & u_a
    	\end{array}
    	,
    \]   
    where the last $ u $ state in the sequence leads only to $ u_a $.   
    
    Now, we provide the details of the transitions. The quantum part has four states and the initial quantum state is $ \mypar{1~~0~~0~~0}^T $.
    
    On the left end-marker symbol $\cent$, $ A $ is in state $ e $ and applies the superoperator with the following operation elements:
    \[   
    	E_{e} = \frac{1}{2}\mymatrix{rrrr}{1 & 0 & 0 & 0 \\0 & 0 & 0 & 0 \\-1 & 0 & 0 & 0 \\0 & 0 & 0 & 0}, ~~
        E_{u} = \frac{1}{2}\mymatrix{rrrr}{1 & 0 & 0 & 0 \\0 & 0 & 0 & 0 \\0 & 0 & 0 & 0 \\0 & 0 & 0 & 0}, ~~
        E_{r} = \frac{1}{2}\mymatrix{rrrr}{1 & 0 & 0 & 0 \\0 & 2 & 0 & 0 \\0 & 0 & 2 & 0 \\0 & 0 & 0 & 2}.    
    \]
    If the outcomes ``$r$'' is observed, $A$ switches to state $ e_r $. If the outcome is ``$e$'', then $ A $ stays in state $e$. Otherwise, $ A $ switches to state $u$.
    
    When $ A $ is in state $ e $ and reads symbol $ a $, it applies the superoperator with the following operational elements:   
    \[
    F_e= \frac{1}{5}\mymatrix{cccc}{1 & 0 & 0 & 0 \\1 & 1 & 0 & 0 \\2 & 0 & 1 & 0 \\2 & 0 & 1 & 1}, ~~
    F_u= \frac{1}{5}\mymatrix{cccc}{1 & 0 & 0 & 0 \\1 & 1 & 0 & 0 \\0 & 0 & 0 & 0 \\2 & 0 & 1 & 1},
   \]
   \[
    F_{r1}= \frac{1}{5}\mymatrix{rrrr}{2 & -1 & 0 & 0 \\2 & 0 & -3 & 0 \\1 & 0 & 0 & -4 \\0 & 0 & 2 & -1},
    F_{r2}= \frac{1}{5}\mymatrix{rrrr}{0 & 4 & 0 & 0 \\0 & 2 & 0 & 0 \\0 & 1 & 2 & 0 \\0 & 1 & -2 & 0},
    F_{r3}= \frac{1}{5}\mymatrix{cccc}{0 & 0 & 1 & 0 \\0 & 0 & 0 & 2 \\0 & 0 & 0 & 1 \\0 & 0 & 0 & 1}.
    \]
    If the outcomes ``r1'', ``r2'', or ``r3'' is observed, $A$ switches to state $ e_r $. If the outcome is ``e'', then $ A $ stays in state $e$. Otherwise, $ A $ switches to state $u$.
    
    When $ A $ is in state $ u $ and reads symbol $ a $, it applies the superoperator with the following operational elements:
     \[
    G_u= \frac{1}{2}\mymatrix{cccc}{1 & 0 & 0 & 0 \\1 & 1 & 0 & 0 \\0 & 0 & 0 & 0 \\0 & 0 & 0 & 1}, ~~
    G_{a1}= \frac{1}{2}\mymatrix{rrrr}{1 & -1 & 0 & 0 \\1 & 0 & 0 & 0 \\0 & 1 & 0 & 0 \\0 & 1 & 0 & 0}, ~~
    G_{a2}= \frac{1}{2}\mymatrix{cccc}{0 & 0 & 2 & 0 \\0 & 0 & 0 & 1 \\0 & 0 & 0 & 1 \\0 & 0 & 0 & 1}.
   \]
   If the outcome ``a1'' or ``a2'' is observed, $A$ switches to state $ u_a $. If the outcome is ``u'', then $ A $ stays in state $u$.
   
  Let us examine the evolution of the (unnormalized) quantum state of a computational path that visits state ``e'' $ m-1 $ times,  and state ``u''  $ n-m+1 $ times after beginning with reading symbol $\cent$, where $ 1 \leq m \leq n $.
   
   After reading $ \cent $, the quantum state is
  \[
   \ket{\widetilde{v_0}} = \frac{1}{2} \mymatrix{r}{1\\0\\-1\\0} \mbox{~~or~~}  \ket{\widetilde{v'_0}} = \frac{1}{2} \mymatrix{r}{1\\0\\0\\0},
   \]
   and the classical state is ``e'' or ``u'', respectively. By supposing to stay in state $ e $, after reading $ (i=1) $ $a$, the quantum state is
   \[
   	\ket{\widetilde{v_1}} = F_e \ket{\tilde{v_0}} =  \frac{1}{2}\frac{1}{5} \myvec{1\\1\\1\\1},
   \]
   and, after reading $ (i>0) $ $a$ symbol(s), the quantum state is
   \[
   	\ket{\widetilde{v_i}} = F_e^i \ket{\widetilde{v_0}} =
     \frac{1}{2}\mypar{\frac{1}{5}}^i \myvec{1\\i\\2i-1\\i^2},  
   \]
   which can be easily verified by induction, i.e.,
   \[
   		\ket{\widetilde{v_{i+1}}} = F_e \ket{\widetilde{v_i}} = 
        \frac{1}{5}\mymatrix{cccc}{1 & 0 & 0 & 0 \\1 & 1 & 0 & 0 \\2 & 0 & 1 & 0 \\2 & 0 & 1 & 1} \frac{1}{2}\mypar{\frac{1}{5}}^i \myvec{1\\i\\2i-1\\i^2} =         
   	\frac{1}{2}\mypar{\frac{1}{5}}^{i+1} \myvec{1\\i+1\\2(i+1)-1\\(i+1)^2}.
   \]
    Thus, after reading $ (m-1) $ $a$ symbol(s), the quantum state is
    \[
    	\ket{\widetilde{v_{m-1}}} = \frac{1}{2}\mypar{\frac{1}{5}}^{m-1} \myvec{1\\m-1\\2(m-1)-1\\(m-1)^2},
    \]
    where $ m>0 $.
    Then, $ A $ switches from state $ e $ to state $ u $ by reading the $ m $-th $ a $, and so, the quantum state is
    \[
    \ket{ \widetilde{ v'_{m} } } = F_u \ket{ \widetilde{ v_{m-1} } } = \frac{1}{2} \mypar{\frac{1}{5}}^{m-1} \frac{1}{2}  \myvec{1\\m\\0\\m^2},
    \]
    where $ m>0 $. (Remember that the quantum state is $ \ket{ \widetilde{ v'_{0} } } $ for $m=0$).
   We can interpret this transition as $ A $ nondeterministically picks $ m $ and also calculates $ m^2 $ and leave the computation to universal to verify whether $ n=m^2 $. The task of the universal part is simple. After applying $ G_u $, the second entry is increased by 1. Thus, after reading the rest of the input, the quantum state is
   \[
   		 \ket{ \widetilde{ v'_{n} } } = \frac{1}{2} \mypar{\frac{1}{5}}^{m-1} \mypar{\frac{1}{2}}^{n-m+1}  \myvec{1\\n\\0\\m^2},
   \]
   where $ n \geq m > 0 $. When $ m=0 $, 
   \[
   		 \ket{ \widetilde{ v'_{n} } } = \frac{1}{2}  \mypar{\frac{1}{2}}^{n}  \myvec{1\\n\\0\\0}.
   \]
   Then, the ultimate decision is given on $\dollar$ symbol by applying the superoperator with the following operation elements:
   \[
   	G'_u= \frac{1}{2}\mymatrix{rrrr}{0 & 1 & 0 & -1 \\0 & 1 & 0 & -1 \\0 & -1 & 0 & 1 \\0 & -1 & 0 & 1}, ~~
    G'_{a}= \frac{1}{2}\mymatrix{rrrr}{2 & 0 & 0 & 0 \\0 & 0 & 2 & 0 \\0 & 0 & 0 & 0 \\0 & 0 & 0 & 0}.
   \]
   If the outcome ``$u$'' is observed, $ A $ stays in state $ u $, and, if the outcome ``$a$'' is observed, $A$ switches to state $ u_a $. 
   
   It is clear that the outcome ``$a$'' is always observed since the corresponding quantum state is
   \[
   	\frac{1}{2} \mypar{\frac{1}{5}}^{m-1} \mypar{\frac{1}{2}}^{n-m+2}  \myvec{2\\0\\0\\0}
    \mbox{~~or~~}
    \frac{1}{2} \mypar{\frac{1}{2}}^{n+1}  \myvec{2\\0\\0\\0},
   \]
     which is observed with probability $ 25^{-m+1} 4^{-n+m-2} > 0 $ or $ 4^{-n-1} >0 $, respectively. On the other hand, the outcome ``$u$'' is observed with non-zero probability only if $ n \neq m^2 $ since the corresponding quantum state is 
     \[
     	\frac{1}{2} \mypar{\frac{1}{5}}^{m-1} \mypar{\frac{1}{2}}^{n-m+2}  \myvec{n-m^2\\n-m^2\\m^2-n\\m^2-n}
        \mbox{~~or~~}
        \frac{1}{2} \mypar{\frac{1}{2}}^{n+1}  \mymatrix{r}{n\\n\\-n\\-n},
     \]
     respectively.
     If $ n=m^2 $, then only the outcome ``$a$'' is observed with non-zero probability and so the input is accepted. Otherwise, the outcome ``$u$'' is also observed and so the input is rejected.
     
     In summary, if $ n $ is a perfect square, then there is non-negative integer $ m $ such that $ n=m^2 $. Hence, $ A $ nondeterministically picks $ m $ and then universally verifies its correctness. (Remark that $ A $ also works correctly on the empty input.) If $ n $ is not a perfect square, then for any non-negative integer, say $ m $, $ n-m^2 \neq 0 $. Thus, any nondeterministic choice of $ A $ is universally falsified.  
\end{proof}

We now show that the emptiness problem for AQFAs over unary alphabets is undecidable.
\begin{thm}
	The emptiness problem for AQFAs over unary alphabets is undecidable.
\end{thm}
\begin{proof} 
	It is known that one-way AQFAs can simulate the computation of a given TM on any input \cite{Yak13A}. In that simulation, the automaton first reads the whole input, and the head arrives at the right end-marker. From that point on, the head never moves, and the automaton  selects the subsequent configurations in the simulation nondeterministically, with each wrong guess  eliminated by the use of universal states. When a halting configuration is obtained, the automaton mimics the corresponding accept/reject decision. The automaton runs forever, but if the TM halts on the input, then the decision can be derived by backing up from a  level of the computation tree that has some finite depth. 

	Let $  M $ be any TM and $ O$ be a one-way AQFA simulating the computation of $ M$ on empty string ($\varepsilon$). Then, we can design a realtime AQFA, say $ R$, that executes some finite steps (depending on the length of the input) of the simulation implemented by $ O$ for $ ({M},\varepsilon) $. Here, $ R$ can execute more steps of the simulation on the longer input string. If ${M}$ accepts $\varepsilon$, then there are some (actually infinitely many) sufficiently long unary strings such that the realtime AQFA accepts these unary inputs.
	
	Therefore, if the emptiness problem for realtime AQFAs over unary alphabet is decidable, then the emptiness problem for TMs is decidable, too. But, this is a contradiction and so the emptiness problem for realtime AQFAs over unary alphabet is undecidable.
\end{proof}

\section{Conclusion}

In this paper, we examine the decidability of emptiness problem for certain generalizations of realtime alternating finite automata: realtime alternating one-counter automaton (A1CA), realtime private and blind alternating finite automata (PAFA and BAFA, respectively), and alternating version of realtime quantum finite automaton (QFA). We also present new alternating algorithms for some well-known unary and binary nonregular languages.

We show that the emptiness problem for A1CA on unary alphabets is undecidable. Then, we show that BAFAs can recognize nonregular languages $ \{ a^nb^n \mid n >0 \} $, $ \twin = \{w2w \mid w \in \{0,1\}^* \} $, $ \upow = \{ a^m \mid m = 2^n \mbox{ for some } n \geq 0 \} $, and $ \usqr = \{ 1^m \mid m = n^2 \mbox{ for some } n \geq 0 \} $. Based on the algorithm for language $ \twin $, we also show that the emptiness problem is undecidable for BAFAs.

Regarding QFAs, we show that the emptiness problem is undecidable both for universal QFAs on general alphabets, and for alternating QFAs with two alternations on unary alphabets. On the other hand, the same problem is decidable for nondeterministic QFAs on general alphabets. We also show that $ \usqr $ is recognized by alternating QFAs with two alternations.

\section*{Acknowledgements}
We thank the anonymous reviewers for their helpful comments. 
Hirvensalo was partially supported by V\"ais\"al\"a Foundation grant 2011-2013.
Yakary{\i}lmaz was partially supported by ERC Advanced Grant MQC, FP7 FET project QALGO, and CAPES with grant 88881.030338/2013-01.

\bibliographystyle{plain}
\bibliography{tcs}

\end{document}